\documentclass[]{aastex631}

\usepackage{subfigure}
\usepackage{amsmath}

\begin{document}

\title{Modeling the X-ray emission of the Boomerang nebula and implication for its potential ultrahigh-energy gamma-ray emission}

\author[0000-0003-4907-6666]{Xiao-Bin Chen}
\affiliation{School of Astronomy and Space Science, Nanjing University, Nanjing 210023, China}
\affiliation{Key laboratory of Modern Astronomy and Astrophysics (Nanjing University), Ministry of Education, Nanjing 210023, China}

\author[0000-0003-3089-3762]{Xuan-Han Liang}
\affiliation{School of Astronomy and Space Science, Nanjing University, Nanjing 210023, China}
\affiliation{Key laboratory of Modern Astronomy and Astrophysics (Nanjing University), Ministry of Education, Nanjing 210023, China}

\author[0000-0003-1576-0961]{Ruo-Yu Liu}
\affiliation{School of Astronomy and Space Science, Nanjing University, Nanjing 210023, China}
\affiliation{Key laboratory of Modern Astronomy and Astrophysics (Nanjing University), Ministry of Education, Nanjing 210023, China}

\author[0000-0002-5881-335X]{Xiang-Yu Wang}
\affiliation{School of Astronomy and Space Science, Nanjing University, Nanjing 210023, China}
\affiliation{Key laboratory of Modern Astronomy and Astrophysics (Nanjing University), Ministry of Education, Nanjing 210023, China}

\correspondingauthor{Ruo-Yu Liu}
\email{ryliu@nju.edu.cn}

\begin{abstract}
The Boomerang nebula is a bright radio and X-ray pulsar wind nebula (PWN) powered by an energetic pulsar, PSR~J2229+6114. It is spatially coincident with one of the brightest ultrahigh-energy (UHE, $\ge 100$\,TeV) gamma-ray sources, LHAASO~J2226+6057.
While X-ray observations have provided radial profiles for both the intensity and photon index of the nebula, previous theoretical studies have not reached an agreement on their physical interpretation, which also lead to different anticipation of the UHE emission from the nebula.
In this work, we model its X-ray emission with a dynamical evolution model of PWN, considering both convective and diffusive transport of electrons.
On the premise of fitting the X-ray intensity and photon index profiles, we find that the magnetic field within the Boomerang nebula is  weak ($\sim 10\mu$G in the core region and diminishing to $1\mu\,G$ at the periphery), which therefore implies a significant contribution to the UHE gamma-ray emission by the inverse Compton (IC) radiation of injected electron/positron pairs. Depending on the particle transport mechanism, the UHE gamma-ray flux contributed by the Boomerang nebula via the IC radiation may constitute about $10-50\%$ of the flux of LHAASO~J2226+6057 at 100\,TeV, and up to 30\% at 500\,TeV. Finally, we compare our results with previous studies and discuss potential hadronic UHE emission from the PWN. In our modeling, most of the spindown luminosity of the pulsar may be transformed into thermal particles or relativistic protons.
\end{abstract}


\section{Introduction} \label{sec:intro}
Pulsar wind nebulae (PWNe) are bubbles of nonthermal electron/positron pairs powered by the rotational energy of pulsars\citep{2006ARA&A..44...17G}. 
Electrons (hereafter, we do not distinguish positrons from electrons unless stated otherwise) are accelerated to ultra-relativistic energies at the termination shock (TS) and give rise to broadband radiations from radio to TeV or even PeV gamma-ray bands, via the synchrotron radiation and the inverse Compton (IC) radiation of the electrons.

\begin{figure}
    \centering
    \includegraphics[width=0.5\linewidth]{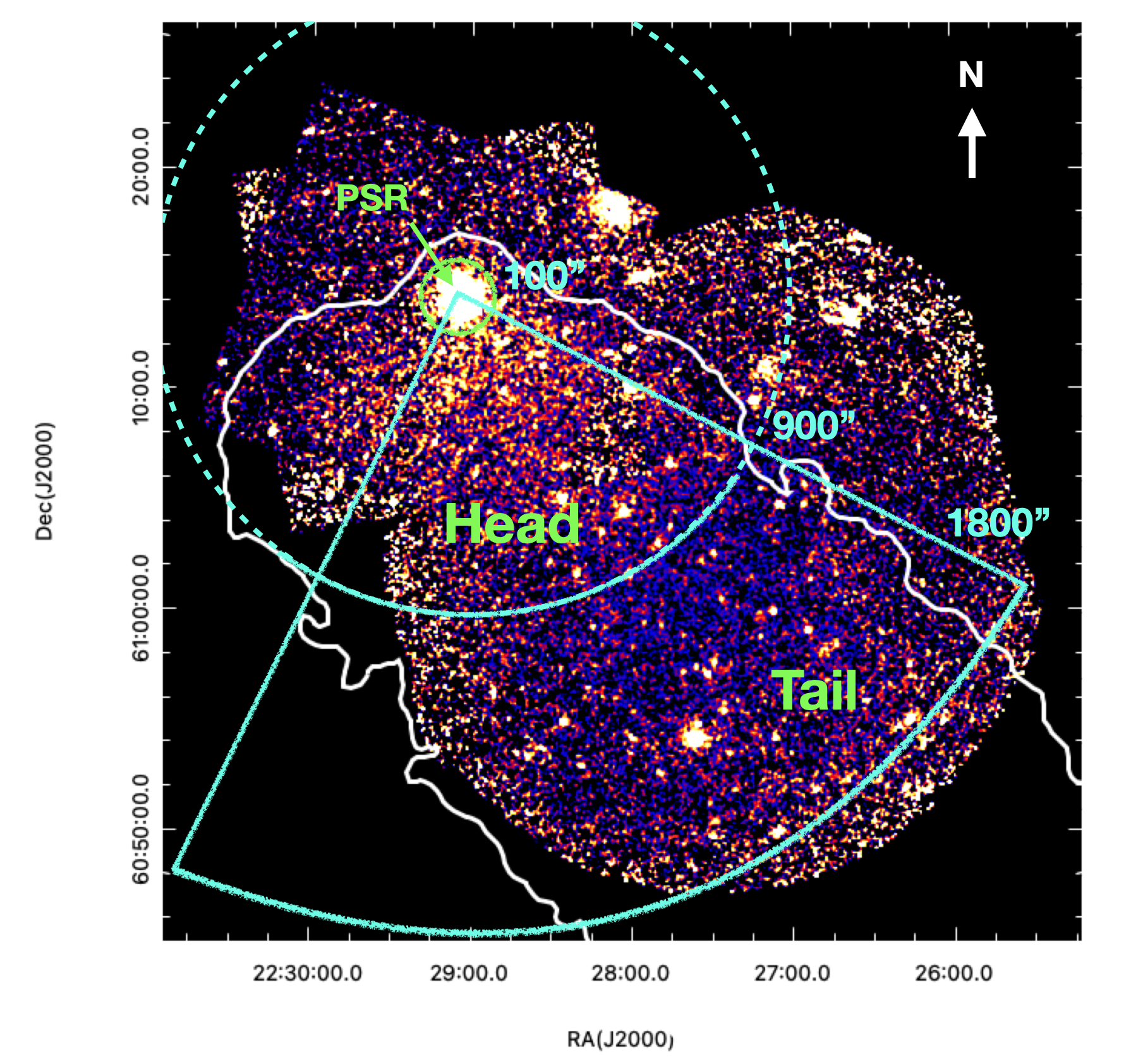}
    \caption{ X-ray Images of the Boomerang nebula. The central-colored part is the combination of the X-ray observation from Chandra and XMM-Newton \citep{2021Innov...200118G}. The green circle has a radius of 100'' centered at the position of PSR J2229+6114. The two cyan arc marks the approximate boundary between the head (900'') and the tail (1800'') of the SNR in the X-ray band. 
    The white curve outlines the 1.4 GHz radio continuum of the SNR \citep{2000AJ....120.3218P}.
    The region roughly represented by a quarter of solid circle extending to the southwest direction is the focus of our model.
    \label{fig:X_ray} }
\end{figure}

The Boomerang nebula is fueled by one of the most powerful pulsars detected so far, i.e., PSR~2229+6114, which boasts a spindown luminosity of $L_s=2.2\times 10^{37}\,$erg/s. 
The pulsar is suggested to be located at about 800 pc away from Earth \footnote{The distance to the Boomerang nebula is uncertain, with estimates ranging from 0.8 kpc \citep{2001ApJ...560..236K} to 7.5 kpc \citep{2009ApJ...706.1331A}, though some evidence suggests it is likely around 2-3 kpc \citep{2001ApJ...547..323H}. On the other hand, the molecular clouds found in the complex seem to have a different spatial distribution, with the location coincident with the tail region in our line of sight based on CO observations \citet{2001ApJ...560..236K}. }, based on the velocity of the HI emission as reported by \citet{2001ApJ...560..236K} and the other parameters we used are listed in Table \ref{tab:psr}. 
It is spatially coincident with the northeast part of an asymmetric radio nebula G106.3+2.7 \citep{1990A&AS...82..113J}, which may be a supernova remnant (SNR) \citep{2000AJ....120.3218P}, although the physical connection between the Boomerang nebula and G106.3+2.7 has not been concretely established. The entire complex consists of a compact, bright ``
head'' in northeast around the pulsar, and an elongated, dimmer ``tail'' extending toward the southwest, as shown by both the nonthermal radio and X-ray emissions \citep{2000AJ....120.3218P, 2020ApJ...897L..34L, 2021Innov...200118G, Fujita21, 2024ApJ...960...75P}, as shown in Fig. \ref{fig:X_ray}.
The intensity of the X-ray emission in the head region decreases with the distance from the pulsar and the spectrum also exhibits a softening trend as the distance increases. The spatial evolution of such properties of the X-ray emission suggests that the emission may originate from the synchrotron radiation of relativistic electrons accelerated in the PWN, which are essentially powered by the pulsar  \citep{2021Innov...200118G}. 

The same population of electrons can also produce gamma-ray emission via the IC radiation. Indeed, the Very Energetic Radiation Imaging Telescope Array System (VERITAS) \citep{2009ApJ...703L...6A,2024ApJ...960...75P}  and the Major Atmospheric Gamma Imaging Cherenkov (MAGIC) \citep{2023A&A...671A..12M} have detected very-high-energy (VHE, $E>0.1\,$TeV) emission
from both the tail region and the head region. Ultra high-energy (UHE, $E>100\,$TeV) emissions have also been detected by High Altitude Water Cherevnkov (HAWC) detector \citep{2020ApJ...896L..29A}, Tibet Air Shower Array (AS$\gamma$) \citep{2021NatAs...5..460T} and Large High Altitude Air Shower Observatory (LHAASO) \citep{2021Natur.594...33C}. 
Particularly, the maximum energy of the LHAASO source, i.e., LHAASO~J2226+6057  or correspondingly 1LHAASO~J2228+6100u in the first LHAASO catalog \citep{2024ApJS..271...25C}, reaches about 500\,TeV. The LHAASO source is extended, spatially overlapping with the Boomerang nebula, and hence part of the UHE emission could in principle originate from the PWN. Note that LHAASO's measurement on the Crab Nebula has demonstrated that young, energetic PWN is capable of accelerating PeV electrons and emitting PeV photons \citep{2021Sci...373..425L}.

A key condition constraining the UHE emission from the PWN is the magnetic field. The ratio between the IC emissivity and the synchrotron emissivity of the same electron population is roughly proportionally to that energy density ratio between the radiation field  and the magnetic field. If the PWN has a strong magnetic field, the IC radiation would be sub-dominated. \citet{2006ApJ...638..225K} suggested a very strong magnetic field of $B=2.6\,$mG inside the Boomerang nebula,  assuming that the break in the radio spectrum is caused by the synchrotron cooling. However, this break could be ascribed to other processes, such as a broken { power-law electron} spectrum formed in the acceleration process \citep{2024ApJ...960...75P}. 

\citet{2022Univ....8..547L} studied this issue by modeling the X-ray intensity and photon index profiles, and obtained a weaker magnetic field of $B=140{\rm \mu G}$ based on a simplified one-zone assumption. This magnetic field is, however, still quite strong (i.e., the magnetic energy density $U_{\rm B}$ is much higher than the background radiation energy density $U_{\rm ph}$), leading to an expected IC flux orders of magnitude lower than the synchrotron radiation. Besides, the synchrotron radiation is very efficient with such a high magnetic field, and therefore only $\sim \bf 10^{-4}$ of the spin-down energy  needs to be converted into the electron energy in order to explain the X-ray flux.
This value is atypical among PWNe, as the value of $\eta_e$ is usually found to be between 0.01 to 1 in previous modeling of other PWNe \citep{chevalierModelXRayLuminosity2000,mayerPredictingXrayFlux2013,2022ApJ...930L...2D,2023ApJ...943...89Z}.
With a dynamical PWN evolution model, \cite{2024ApJ...960...75P} explained the X-ray spectrum of the nebula assuming a larger distance of 7.5\,kpc for the pulsar. 
However, \cite{2024ApJ...960...75P} did not take into account the radial profiles of X-ray intensity and photon index in the modeling, and hence it is not clear whether the spatial evolution of the X-ray emission can be reproduced with the employed parameters.

In this work, we aim to reproduce both the spectrum and radial profiles of the X-ray emission in the head region with a dynamic PWN model. We will consider various particle transport scenarios, including convection-dominated, convection-diffusion, and diffusion-dominated  transports. We constrain critical PWN parameters such as the magnetic field, and estimate the UHE emission of electrons accelerated in the nebula. The rest of the paper is organized as follows: the dynamic PWN model is briefly reviewed in Section \ref{sec:model}. We present the fitting result for the SED and radial profiles in Section \ref{sec:solutions}. In Section \ref{sec:TeV}, we estimate the UHE gamma-ray emission generated by IC radiation of electrons. We discuss the results in Section \ref{sec:Discu} and conclusions are given in Section \ref{sec:Conclu}.

\begin{table}
    \centering
    \begin{tabular}{lc}
    \hline
        parameter (unit)       & value \\
    \hline
        $P\rm(s)$ [1]              & 0.0516 \\
        $P_0\rm(s)$            & 0.04\\
        $\dot{P}\rm(s~s^{-1})$ [1] & $7.827 \times 10^{-14}$\\
        $L_s \rm (erg~s^{-1})$ [1] & $2.2 \times 10^{37}$\\
        $T_{\rm age}$(yr)      & 4200  \\
        $d\rm(pc)$  [2]           & 800 \\
    \hline
    \end{tabular}
    \caption{The basic properties of PSR~J2229+6114. References: [1] \cite{2001ApJ...552L.125H}; [2] \cite{2001ApJ...560..236K}.}
    \label{tab:psr}
\end{table}

\section{PWN Model and Radiation} \label{sec:model}

In this section, we describe the model to calculate the transportation of the relativistic electrons in the Boomerang nebula and the radiation from these electrons. For simplicity, three assumptions are made:
(1) in the PWN, a constant fraction $\eta_e$ of the spindown power of the pulsar is injected into the PWN in the form of relativistic electrons during the evolution;
(2) particle injection occurs at the termination shock (angular size $\sim 10''$ according to  \citealt{2004ApJ...601..479N}, corresponding to $R_{\rm ts} = 0.035$\,pc at a nominal distance of 800\,pc);
(3) an approximation of spherical symmetry of system and the dynamical evolution  model proposed by \cite{2011MNRAS.410..381B} is adopted. 
For assumption (3), we note that the Boomerang nebula actually does not show a symmetric morphology. Its northeast part seems to be interacting with surrounding atomic gas \citep{2001ApJ...560..236K} and hence appear compact, while its southwest part is more extended. The X-ray profile measurement is also mainly focused on a sector in the southwest direction \citep{2021Innov...200118G}. From observations we did not see a clear dependence of the intensity on the azimuthal angle. In other words, the profile only has a radial dependence, so an approximation of spherical symmetry in the modeling is acceptable. 

\subsection{Model Descriptions}
The energy input into the PWN is proportional to the spindown power of the pulsar $L_s$. Assuming the pulsar as a magnetic dipole, the temporal evolution of the spindown power is usually described as
\begin{equation}
L_s(t)=\frac{L_0}{(1+t/\tau_0)^2},
\end{equation}
where $L_0$ is the initial spindown luminosity and $\tau_0$ is the initial spindown timescale. $\tau_0$ can be calculated as $\tau_0=(P/2\dot{P})(P_0/P)^2$, $P$, $\dot{P}$ and $P_0$ being the current rotational period, first derivative of the period and the initial rotational period of the pulsar. The initial rotational period $P_0$ is unknown and hence depends on assumption.   
In general, the dynamical evolution of a PWN could be roughly divided into three stages, which define the expansion behaviour of the PWN according to the age of the system $t_{\rm age}$ in comparison to the initial spindown time $\tau_0$ and the supernova remnant reverse-shock passage time $t_{\rm rs}$ \citep{Reynolds84}, which is typically several thousands of years and the detailed value depends on the ambient gaseous environment. The true age of the pulsar can be calculated by $t_{\rm age}=(P/2\dot{P})[1-(P_0/P)^2]$. The current rotational period $P$ of PSR~2229+6114 closes to our estimated initial period, so the assumed initial rotational period $P_0$ have a non-negligible influence on the $t_{\rm age}$ and consequently the evolutionary stage of the PWN. When the reverse shock reaches the PWN, it may crush and compress the PWN. In the northeast part of the nebula, the presence of the dense atomic gas in proximity of the pulsar may lead to formation of a strong reverse shock at early time, and may provide an explanation to the compactness of the nebula in the northeast part. On the the hand, there is a gas cavity in the southwest part of the pulsar, and hence the reverse shock may not form (at least we do not expect formation of a strong reverse shock) or has not reached the PWN. Since we focus on the X-ray radial profiles measured toward the southwest part of the nebula, we choose a value of $P_0=40\,$ms, corresponding to a value of $t_{\rm age}\simeq 4200\,$yr, which is smaller than the typical reverse-shock passage time $t_{\rm RS}$ given a low density of the ambient medium ($t_{\rm RS} \simeq  1.05 \tau_0$ following Eq.(29) in \citealt{2023MNRAS.520.2451B}).  With such a $P_0$, we may also obtain $\tau_0 \simeq 6000\,$yr. As $t_{\rm age}$ is smaller than both $t_{\rm RS}$ and $\tau_0$, the PWN is currently in the phase of free expansion.

At this stage, the time evolution of the outer radius of the PWN $R_{\rm pwn}$ is then described as \cite{2001A&A...380..309V} by 
\begin{equation}
    R_{\rm pwn}(t) = 0.839 \left(\frac{L_0 t }{E_{\rm sn}}\right)^{1/5} \sqrt{\frac{10}{3}\frac{E_{\rm sn}}{M_{\mathrm{ej}}}} ~t
\end{equation}
where $E_{\rm sn}$ is the total mechanical energy of the supernova explosion energy, setting $E_{\rm sn}= 10^{51}$~erg, and $M_{\mathrm{ej}}$ is ejecta mass, which is set to 5\,$M_\odot$.
According to PWN Magneto-hydrodynamic (MHD) models \citep{2006A&A...454..393B}, the majority of the spin-down power is released in a relativistic particle wind.
For the energy spectrum of electrons freshly injected into the nebula we assume the following power-law shape:
\begin{equation}
    Q_{\rm inj}(\gamma,t) = Q_0(t) \times \gamma^{-\alpha}, \ 
    \gamma_{\rm min} < \gamma < \gamma_{\rm max}
\end{equation}
with a power-law index $\alpha$.
In the above equations, $Q_0(t)$ and $\gamma$ are the normalization coefficient as well as the Lorentz factor of electrons.
$\gamma_{\rm min}$ and $\gamma_{\rm max}$ are the minimum and maximum Lorentz factors of electrons. We set $\gamma_{\rm min} = 2000$ and $Q_0(t)$ is estimated by
\begin{equation}
    \eta_e L(t)=\int_{\gamma_{\rm min}}^{\gamma_{\rm max}} 
    \gamma m_e c^2 Q_\mathrm{inj}(\gamma,t)d\gamma.
\end{equation}

In a spherically symmetric system, the transport of particles with number density $n = n(r, \gamma, t)$ within the nebula can be written as the Fokker–Planck equation \citep{1965P&SS...13....9P}: 
\begin{equation} \label{eq:dndt}
\frac{\partial n}{\partial t} = 
D \frac{\partial^2n}{\partial r^2} + 
\left[\frac1{r^2}\frac\partial{\partial r}(r^2D)-V\right]\frac{\partial n}{\partial r} - 
\frac1{r^2}\frac\partial{\partial r}[r^2V]n + 
\frac\partial{\partial\gamma}[\dot{\gamma}n]+Q_{\rm inj}.
\end{equation}
In the above equation, $\dot{\gamma}$ is the summation of particle energy losses, including the adiabatic expansion loss, synchrotron radiation energy loss and IC scattering energy loss, $D$ denotes the diffusion coefficient, $V$ is the bulk velocity of electrons i.e. the convection velocity, and the function $Q_{\rm inj}$ is the distribution of particles injected from the termination shock.
On the other hand, to simulate the particles escaping from the PWN, a free escape condition is imposed at the outer boundary: $n(r=R_{\rm pwn})=0$ \citep{2013ApJ...765...30V}.

MHD simulations show a complicated flow and magnetic field structure in PWN (e.g. \citealt{1984ApJ...283..694K}). It is further assumed that the ideal MHD limit, allowing one to relate the radial profiles of the velocity and magnetic field: 
$\nabla \times \boldsymbol{V} \times \boldsymbol{B} = 0$.
Without loss of generality, $V(r)\propto r^{-\beta}$ is assumed.
Steady-state MHD simulations \citep{1984ApJ...283..694K} find that the radial velocity profile depends on the ratio of electromagnetic to particle energy, $\sigma$. When $\sigma=1$, the velocity remains almost independent of $r$. When $\sigma=0.01$, the velocity falls off as $V(r)\propto r^{-2}$ at the nebula inner region and approaches a constant as the nebula radius increases to a hundred times of the termination shock radius. For the current model $0<\beta \le 2$ is chosen. 
With the chosen flow and magnetic field structure, we get $ VBr={\rm constant}=V_0B_0r_0$.
Then the bulk velocity can be expressed as 
\begin{equation}
    V(r)=V_0\Bigg(\frac r{R_{\mathrm{pwn}}(t)}\Bigg)^{-\beta},
\end{equation}
where $V_0$ is the convective flow downstream of the pulsar wind, estimated to be $R_{\rm pwn}(T_{\rm age})/T_{\rm age} \simeq$ 140\,km/s \citep{2022ApJ...926....7P}.
The radial profile of the magnetic field inside the nebula can be given by
\begin{equation}
    B(r,t)=B_0(t)\Bigg(\frac{r}{R_{\mathrm{ts}}(t)}\Bigg)^{\beta-1},
\end{equation}
where $B_0(t)$ is the magnetic field at the termination shock, calculated by integration of magnetic energy in space $W_B(t) = \eta_B \int L(t)dt =\int B^2(r,t)r^2dr/2$.
$\eta_B$ describes the fraction of the magnetic energy converted from the spin-down luminosity. 
In an expanding system, the magnetic energy is balanced by the injection rate of electromagnetic energy and the adiabatic loss \citep{1973ApJ...186..249P}.

The maximum energy of accelerated electrons are mainly determined by two processes, confinement and cooling\citep{2022ApJ...930L...2D}. The former constraint can be related to the potential drop of the pulsar\cite{1992ApJ...396..161D}, and we follow the previous studies \citep[e.g.,][]{2022ApJ...926....7P}
\begin{equation} \label{eq:gmax}
\gamma_{\max}\approx\frac{4\epsilon e}{m_ec^2}\sqrt{\frac{L(t)}c\frac\sigma{1+\sigma}}=\frac{4\epsilon e}{m_ec^2}\sqrt{\eta_\mathrm{B}\frac{L(t)}c},
\end{equation}
where $e$ is the electron charge, and $\eta_\mathrm{B}=\sigma/(1+\sigma)$. 
The particle’s Larmor radius, $r_L = E/(ZeB_{\rm acc})$, \textbf{where $Z=1$}, should not be larger than $R_{\rm ts}$. $B_{\rm acc}$ is the magnetic field of the acceleration zone, which we consider as the magnetic field of the termination shock, i.e., $B_{\rm acc}=(4/R_{\rm ts})\sqrt{\eta_BL(t)/c}$. Hence $\epsilon = r_L/R_{\rm ts}$ is the ratio of electron's Larmor radius to the pulsar wind termination shock radius with $0<\epsilon \le 1$, which can be also regarded as the acceleration efficiency. The constraint from radiative cooling reads \citep{2021Sci...373..425L}
\begin{equation}\label{eq:gmax2}
    \gamma_{\max}=1.1\times 10^{10}\epsilon^{1/2}\left(\frac{B_{\rm acc}}{100\mu \rm G}\right)^{-1/2},
\end{equation}
considering the synchrotron radiation as the dominant energy loss process. The maximum electron energy is equal to the minimum between Eq.~(\ref{eq:gmax}) and Eq.~(\ref{eq:gmax2}).

The particle diffusion coefficient, which is related to the magnetic field as $\propto 1/B(r,t)$ and particle energy as $\propto E^{1/3}$ (for a typical Kolmogorov turbulence diffusion e.g., \citealt{1941DoSSR..30..301K}), can be expressed as 
\begin{equation}\label{eq:D}
    D(r,E,t)=D_0 {\left(\frac r{R_{\rm ts}(t)}\right)}^{1-\beta} \left(\frac{E}{100 \rm TeV}\right)^{1/3}
\end{equation}
where $D_0$ is diffusion coefficient for 100 TeV at the termination shock.

\subsection{Radiation}

Relativistic electrons produce photon emission via synchrotron radiation in the interstellar magnetic field and IC scattering off interstellar background photons.
The target photon fields for IC scattering include cosmic microwave background (CMB), starlight, and infrared photons. 
The latter two can be represented by greybody radiation.
The energy densities of the starlight and infrared fields are assumed to be 0.1\,eV~cm$^{-3}$ and 0.1\,eV~cm$^{-3}$, respectively. 
The temperatures at the spectral peaks are 5000\,K for the starlight field component and 70\,K for the infrared \citep{2005ICRC....4...77P}.

Following the method given by \cite{2019ApJ...875..149L}, the predicted intensity of synchrotron radiation at an angular distance $\theta$ from the pulsar’s direction can be calculated by integrating over the contribution of electrons along the line of sight of that direction:
\begin{equation}
I_{\rm syn/ic}(\epsilon,\theta)=\frac1{4\pi}\int\mathcal{F}_{\rm syn/ic}\{N[E,r(\theta,l)],U_{\rm B/ph} \}\mathrm{d}l,
\end{equation}
where $\mathcal{F}_{\rm syn/ic}$ are operators calculating the differential spectrum of synchrotron radiation and IC radiation of electrons $F(\epsilon)$ following the formulae in \cite{1979rpa..book.....R}, given the electron density, the magnetic field, or the background photon field.
$U_{\rm B/ph}$ represents the energy density of a magnetic field or background photon field. The total flux can be obtained by integrating over the solid angle around the pulsar.
$N(E,r)$ is the differential number density of electrons at different positions.

\citet{2021Innov...200118G} suggested that SNR~G106.3+2.7 are accelerating high-energy electrons and responsible for the flat X-ray intensity profile in the tail region. Due to the projection effect, a subdominant fraction of the emission from the head region may also arise from the SNR shock. We assume the flat X-ray intensity profile generated by the SNR extends to the head region, following the treatment in \citet{2022Univ....8..547L}.

\section{Solutions to the transport equation}\label{sec:solutions}

In this section, we have conducted a detailed study on the various transport mechanisms of electrons, considering three different scenarios for electron transport: convection-dominated, convection-diffusion, and diffusion-dominated.
To comprehend the impact of convection and diffusion on particle spectra, we present the formulas for their timescale as $\tau_{\rm con} = \int_{r_0}^r 1/{V(r)}{dr}$ and $\tau_{\rm diff} = r^2/D$.

\begin{table}
\renewcommand\arraystretch{1.5}
\setlength\tabcolsep{10pt}
    \centering
    \begin{tabular}{lcccccc}
    \hline
        parameter & $ \log \eta_B$  & $\log \eta_e$ & $\alpha$ & $\beta$ & $\epsilon$ &  $B_0(t=\rm now)$\\
    \hline    
        Convection-dominated & ${-1.83}^{+0.15}_{-0.11}$ & ${-1.33}^{+0.25}_{-0.20}$ & ${2.00}^{+0.10}_{-0.10}$ & ${0.23}^{+0.02}_{-0.02}$ & ${0.91}^{+0.07}_{-0.13}$ & ${7.4}^{+3.1}_{-1.9} \rm \mu G$\\
        Convection-diffusion & ${-2.07}^{+0.26}_{-0.24}$ & ${-1.61}^{+0.18}_{-0.17}$ & ${1.69}^{+0.14}_{-0.15}$ & ${0.28}^{+0.03}_{-0.03}$  & ${0.57}^{+0.19}_{-0.14}$ & ${7.3}^{+5.9}_{-3.1} \rm \mu G$\\
        Diffusion-dominated & ${-1.88}^{+0.40}_{-0.66}$ & ${-1.78}^{+0.68}_{-0.33}$ & ${1.25}^{+0.30}_{-0.30}$ & ${0.31}^{+0.03}_{-0.03}$  & ${0.23}^{+0.35}_{-0.09}$ & $ {10.2}^{+15.4}_{-7.9} \rm \mu G$ \\
    \hline
    \end{tabular}
    \caption{First row: model parameters; second to fourth lines: best-fit values and 1$\sigma$ uncertainties for three scenarios. The seventh column provides the current magnetic field near the TS plane under the best fit scenario.}
    \label{tab:MCMC}
\end{table}

\subsection{Convection-dominated} \label{subsec:con_only}
In the convection-dominated scenario, $\tau_{\rm con} \ll \tau_{\rm diff}$, we simply ignore the term related to diffusion (i.e., $D=0$ is adopted) in Equation~(\ref{eq:dndt}). This scenario may occur if the magnetic field in the PWN is toroidal, such as observed by IXPE in the Crab Nebula \citep{2023NatAs...7..602B}, so that the diffusion along the radial direction will be largely inhibited.

\begin{figure}
    \centering
    \includegraphics[width=0.45\linewidth]{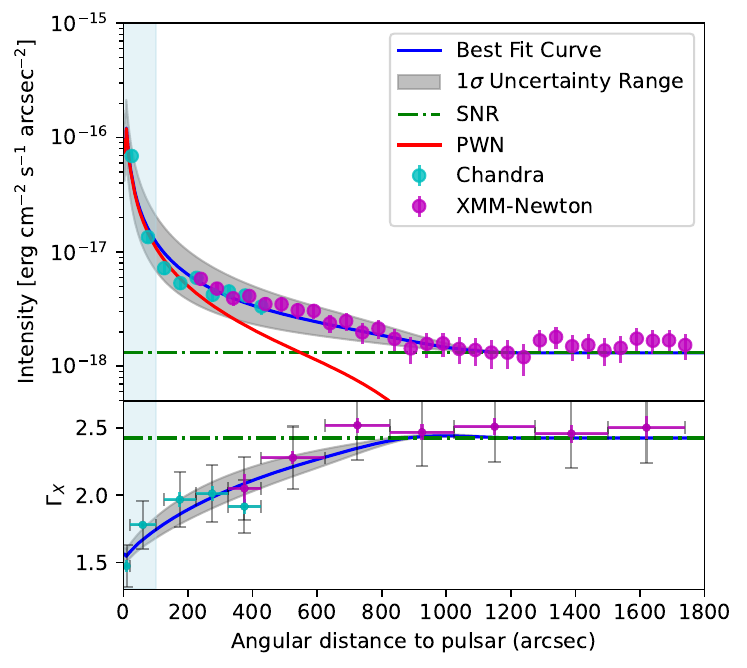}
    \includegraphics[width=0.43\linewidth]{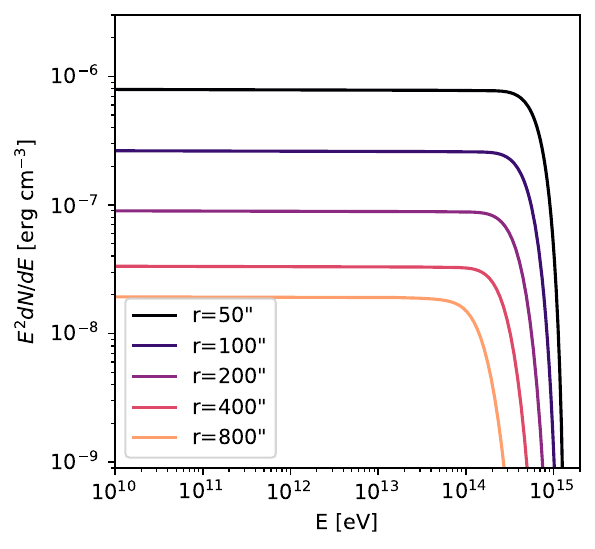}
    \caption{{\bf Left panel}: the fitting result of the radial profiles of X-ray surface brightness and photon index for convection-only scenario. 
    The optimal fitting values for MCMC are: $\eta_B = 0.015,~\eta_e = 0.046,~\alpha=2.001,~\beta=0.231,~\epsilon= 0.91$.
    The red line represents the PWN electron component, the green dash-dotted line represents the SNR electron component, and the blue solid line with 1$\sigma$ error band (gray) is the sum of the two.
    Cyan and magenta data points are from Chandra and XMM-Newton, respectively, taken from \cite{2021Innov...200118G}.
    Displayed statistical and system errors, with a system error of 10\% and indicated in black, due to calibration uncertainties of instruments \citep{2015A&A...575A..30S}.
    {\bf Right panel}: local electron spectra at different radius for the convection-dominated scenario.}
    \label{fig:D1e_3}
\end{figure}

Including the effect of synchrotron losses lead to the electron spectrum's cutoff ${E}_{\rm cut,syn}$, which shifts to lower energies with increasing $r$. 
Because synchrotron energy loss dominates for UHE electrons, the energy ${E}_{\rm cut,syn}$, where the synchrotron break should appear, can be estimated by equating the convection time scale $\tau_{\rm con}$ and synchronous cooling time scale $\tau_{\rm syn} \sim {8.4\times10^3}/(B_{\mu \rm G}^2E_{\rm TeV})\rm ~kyr$.
The maximum injection energy of electrons will not affect the results in this scenario, if $E_{\rm cut, syn}<E_{\rm max}$. 

We use Markov-chain Monte Carlo method (MCMC) to find the best-fit parameters with the Python package {\it emcee} \citep{2013PASP..125..306F}.
The results are listed in Table \ref{tab:MCMC}.
Then represent the scenario corresponding to the best fit values given by MCMC and the 1-sigma error band in Figure~\ref{fig:D1e_3}, fitting radial profiles of X-ray surface brightness and photon index of the PWN-SNR complex at 1-7 keV. 
The y-axis of the upper panel in Figure~\ref{fig:D1e_3} is intensity and that of the lower one is photon index, resembling Figure~3A from \cite{2021Innov...200118G}. 
The light blue shadow specifically highlights the region within 100'', where the brightness is significantly higher than the outer part. Our best fitting result also reproduces the decrease in X-ray intensity.
It is clear that the radial variations of the X-rays' photon index and surface brightness  detected by the XMM-Newton and Chandra can be reproduced in the frame of our model.

When the propagation of electrons depends only on energy-independent convection, spectral shape of electrons is maintained during propagation (if cooling is neglected), and thus the index of the injected electron spectrum need not be very hard ( i.e., fine with $\alpha \gtrsim 2$). 
Due to the radiative cooling and adiabatic cooling of UHE electrons, $E_{\rm max}$ of propagated electron spectrum varies with the distance $r$.
The right panel of Figure~\ref{fig:D1e_3} shows the spatial evolution of electron spectrum at different radius.
Since $E_{\rm max}$ is significantly related to the peak of synchrotron radiation, a noticeable softening of the spectrum can be reproduced in X-ray observations from 1-7\,keV given an spatial distribution of the magnetic field, .
According to the best-fit result, the magnetic field has the values $B_{\rm r=10''} \approx 7\mu$G and $B_{\rm r=900''} \approx 2 \mu$G. 
For the electrons maximum energy, we obtain $E_{\rm max,~r=10''} \approx 1.1$\,PeV and $E_{\rm max,~r=900''} \approx 700$\,TeV.

\subsection{Convection-diffusion}

\begin{figure}
    \centering
    \includegraphics[width=0.45\linewidth]{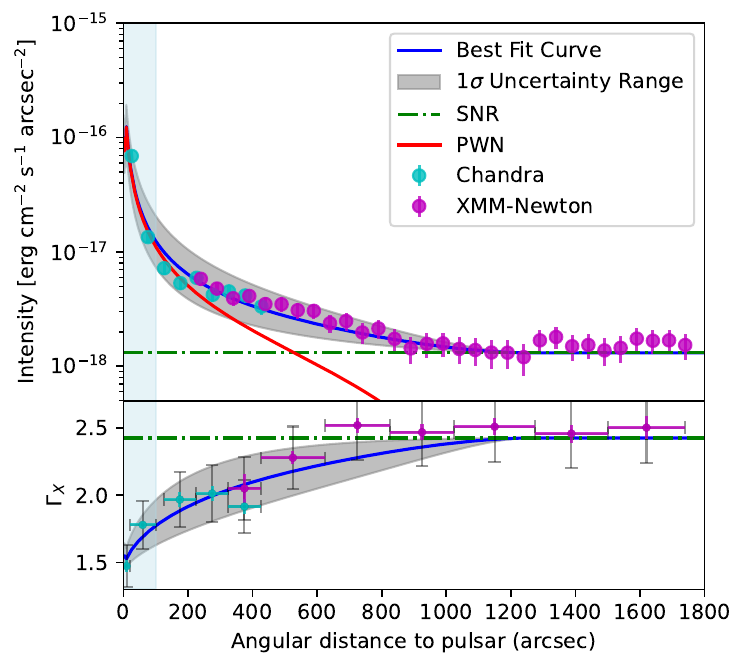}
    \includegraphics[width=0.43\linewidth]{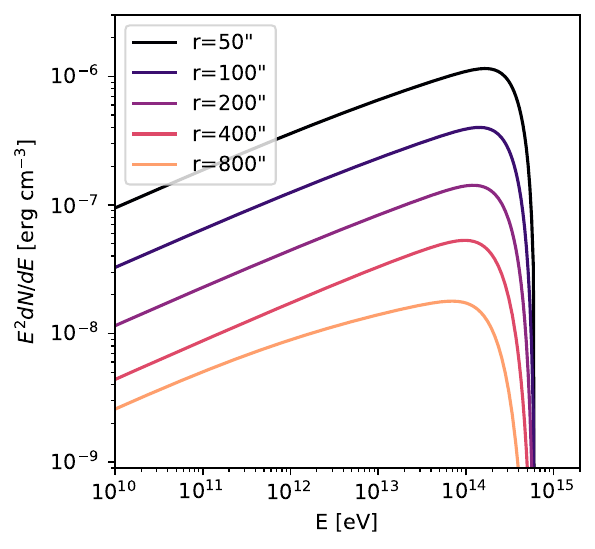}
    \caption{{\bf Left panel}: fitting result of the radial profiles of X-ray surface brightness and photon index for convection-diffusion scenario. 
    The optimal fitting values for MCMC are: $\eta_B = 8.458\times10^{-3},~\eta_e = 0.0246,~\alpha= 1.692,~\beta=0.2836,~\epsilon= 0.570$.
    {\bf Right panel}: local electron spectra at different radius for the convection-diffusion scenario.
    Legend same as Figure~\ref{fig:D1e_3}.}
    \label{fig:D1e_1}
\end{figure}

Diffusion of particles may also play an important role on the electron distribution, if some turbulence is present in the toroidal magnetic field. More quantitatively, if timescales between convection and diffusion, i.e., $\tau_{\rm con} \sim \tau_{\rm diff}$, are comparable, we cannot neglect the term related to $D$ in Eq.~(\ref{eq:dndt}).
This scenario corresponds to  a {\bf diffusion coefficient of ${D_0} \rm (100TeV)\sim 10^{26} cm^2~s^{-1}$ (for $\beta = 0.3$)} for the energy range of interest. 
Note that,since the diffusion coefficient is usually energy dependent, such as $D \propto E^{1/3}$ for the Kolmogorov-type turbulence, the particle transport may be still dominated by energy-independent convection at low energies. Figure~\ref{fig:D1e_1} shows the fitting results for the convection-diffusion scenario.

In the context of a convection-diffusion system, the evolution of spectra can be categorized into two distinct regimes. At lower energies, convection and adiabatic losses play a significant role, thereby preserving the shape of the source spectrum during propagation. With increasing electron energy, diffusion becomes the predominant factor, resulting in a softening in the electrons spectral due to energy-dependent diffusion (although not evident in the figure), as shown in the right panel of Figure~\ref{fig:D1e_1}. At the large distance of $r=800"$, the spectrum presents a more apparent softening at the high energies because the implementation of the free-escape (Dirichlet) condition at the system's outer boundary, akin to the findings reported in \cite{2013ApJ...765...30V}.
And according to the best-fit result, for the electrons maximum energy, we obtain $E_{\rm max,~r=10''} \approx 950$\,TeV and $E_{\rm max,~r=900''} \approx 600$\,TeV.

Compared to the convection-dominated scenario, diffusion reduces the propagation time of particles through the system, thereby relaxing synchrotron losses suffered by the particles. 
The evolution of the high-energy cutoff in the spectrum caused by synchronous energy loss with radius is much weaker than the previous scenario.
Therefore, the reason for the softening of X-ray photon index with radius can no longer be solely explained with the decline of the high-energy spectral cutoff. A decrease in the magnetic field with radius also contributes to the observed radial softening of the X-ray photon index.
At the same time, we can also find that the required injected electron spectrum in this case is harder due to the influence of diffusion.

\subsection{Diffusion-dominated}

If the diffusion coefficient is as high as $D \rm (100TeV) \sim 10^{27} cm^2~s^{-1}$, i.e. $\tau_{\rm con} \gg \tau_{\rm diff}$ for UHE electrons, diffusion dominates the transport process of these UHE electrons, and the contribution of convection can be ignored. 
According to previous studies \citep{2009ApJ...693.1275A, 2015PhRvD..92h3003P}, the diffusion equation may encounter the problem of unrealistic superluminal propagation within a distance of $r<D/c$ from the particle injection point. We therefore follow the treatment of \citet{2015PhRvD..92h3003P, 2022Univ....8..547L} to solve the particle transport in this scenario although the superluminal propagation only occurs at very small scales with the considered value of the diffusion coefficient.

\begin{figure}
    \centering
    \includegraphics[width=0.45\linewidth]{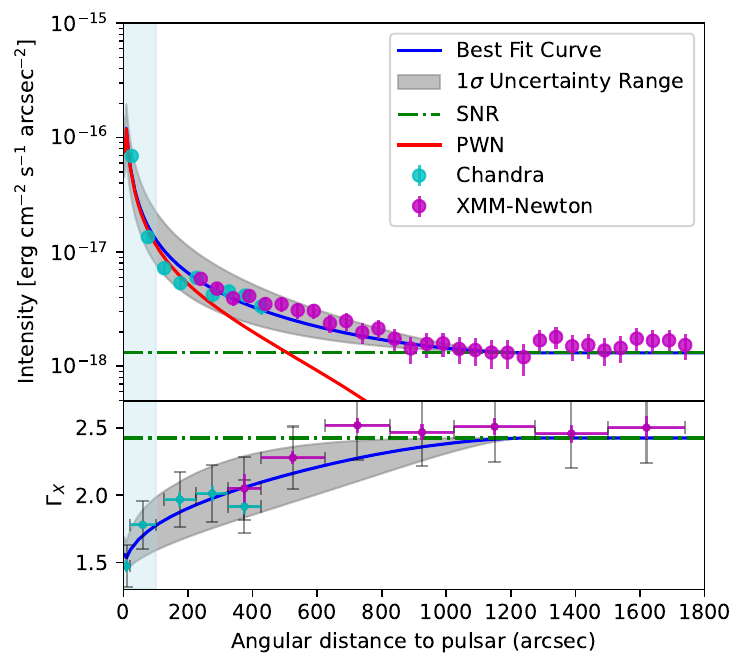}
    \includegraphics[width=0.43\linewidth]{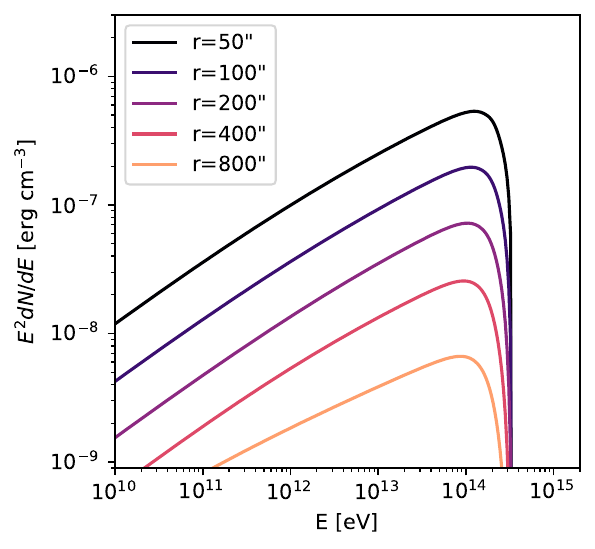}
    \caption{{\bf Left panel}: fitting result of the radial profiles of X-ray surface brightness and photon index for Diffusion-dominated scenario. 
    The optimal fitting values for MCMC are: $\eta_B = 0.0223,~\eta_e = 0.0117,~\alpha= 1.476,~\beta=0.3146,~\epsilon= 0.1893$.
    {\bf Right panel}: local electron spectra at different radius for the diffusion-dominated scenario.
    Legend same as Figure~\ref{fig:D1e_3}.}
    \label{fig:D1}
\end{figure}

Figure~\ref{fig:D1} shows the fitting results for the diffusion-dominated scenario. 
Different from convection-only scenario, electrons do not cool efficiently, because with a relatively fast diffusion considered in this
scenario, electrons diffuse quickly to larger radius, where the magnetic field is weaker.
As shown in the right panel of the figure \ref{fig:D1e_3},
the gradual softening of the photon spectrum is mainly determined by the spatial evolution of the magnetic field. The magnetic field strength is smaller at a larger distance from the source center, so the peak of the synchrotron radiation moves toward lower frequency at a larger distance. As a result, the value of $\beta$ (i.e., the spatial dependence of the magnetic field) is the highest among all three scenarios, as can be seen from Table~\ref{tab:MCMC}. From the table, we also see that the injection spectral index in this scenario is harder than the previous two scenarios. This is because the propagated electron spectrum is softened by $1/3$ (for a typical Kolmogorov-type turbulence) in this scenario, and hence it requires a harder injection electron spectrum to reproduce the measured X-ray photon index. 
And according to the best-fit result, for the electrons maximum energy, we obtain $E_{\rm max,~r=10''} \approx 500$\,TeV and $E_{\rm max,~r=900''} \approx 330$\,TeV.

\section{Contribution to the gamma-ray band}\label{sec:TeV}

The gamma-ray observations of the Boomerang region cover a wide energy range from GeV to PeV \citep{2009ApJ...703L...6A, 2020ApJ...896L..29A, 2021Natur.594...33C, 2023A&A...671A..12M, 2024ApJ...960...75P}.
In particular, \cite{2021Natur.594...33C} report the LHAASO J2226+6057 with the maximum photon energy of over 500\,TeV.
Although the centre of the LHAASO source deviates from the pulsar’s position by about $0.3^\circ$, the source partially covers the Boomerang nebula, and hence the latter may be a contributor.

\begin{figure}
    \centering
    \subfigure[TeV emission for convection-dominated.]{\includegraphics[width=0.32\textwidth]{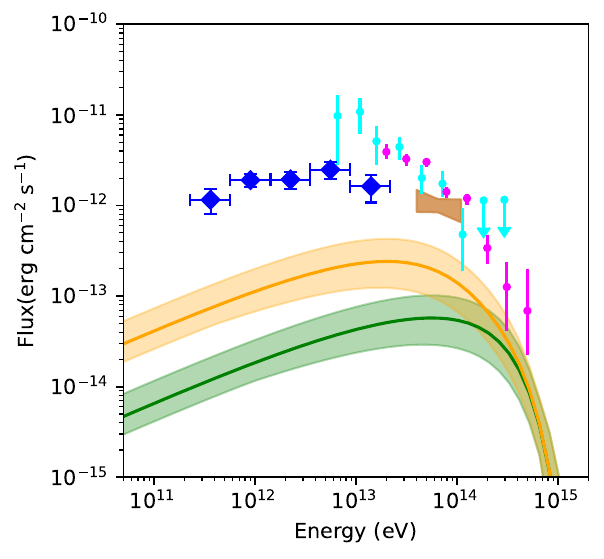}\label{fig:subfig_a}}
    \hfill
    \subfigure[TeV emission for convection-diffusion.]{\includegraphics[width=0.32\textwidth]{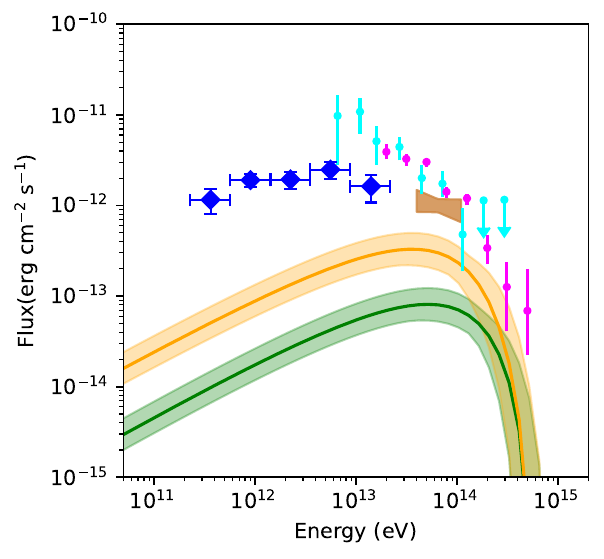}\label{fig:subfig_b}}
    \hfill
    \subfigure[TeV emission for diffusion-dominated.]{\includegraphics[width=0.32\textwidth]{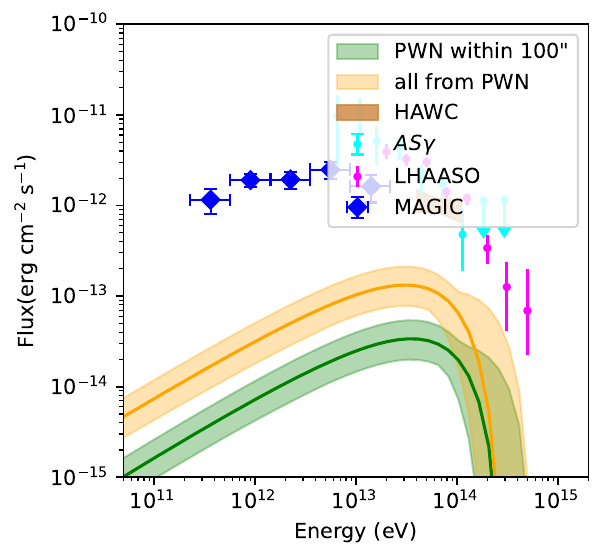}\label{fig:subfig_c}}
    \caption{The multi-wavelength spectral energy distribution of PWN Boomerang for three scenarios. The green line represents a circular area with a radius of 100" around the pulsar, and the orange line depicts a southwest extension structure stretching 900'', with the width of each line indicating the allowed error range of 1 sigma.
    The blue, cyan and magenta TeV data points are from MAGIC \citep{2023A&A...671A..12M}, AS$\gamma$ \citep{2021NatAs...5..460T} and LHAASO \citep{2021Sci...373..425L}, respectively. Brown butterfly plot is from HAWC \citep{2020ApJ...896L..29A}. }\label{fig:SED_TeV}
\end{figure}

Based on our modeling of the X-ray data, we can calculate the IC radiation of electrons. The generated SEDs under three transport scenarios are shown as the solid curves in Figure~\ref{fig:SED_TeV}. 
Following the treatment of \cite{2021Innov...200118G}, we divide the Boomerang nebula into two regions for research. One is a circular area with a radius of 100" around the pulsar, and the other is a more extended area including the head region of SNR~G106.3+2.7, roughly resembling a quarter circle stretching out to 900" from the pulsar. In the Figure~\ref{fig:SED_TeV}, the green curve and the orange curve represent, respectively, the expected IC flux within 100" and 900". The shaded region show the corresponding $1\sigma$ uncertainties of the model prediction. 
It can be seen that, the IC radiation above 100\,TeV mainly arise from electrons within 100". This is due to rapid cooling of UHE electrons so that they cannot be transported to distant area. In principle, the Boomerang nebula may contribute to a non-negligible fraction of the highest-energy emission of LHASO~J2226+6057.
Depending on the particle transport scenario, the UHE gamma-ray flux contributed by the Boomerang nebula via the IC radiation may constitute about $10-50\%$ of the flux of LHAASO~J2226+6057 at 100\,TeV, and up to 30\% at 500\,TeV.

\cite{2023A&A...671A..12M} also reported the energy spectra in energy regimes from 0.2 TeV to 20 TeV of the head and tail regions using MAGIC telescopes. The flux predicted by our model in this energy range is 1-2 orders of magnitude lower than the observed one. This is consistent with the conclusion drawn by \cite{2023A&A...671A..12M}, which suggests that in this wavelength band, the radiation may not originate from electron radiation.

\section{Discussion} \label{sec:Discu}

\subsection{Comparison with  \cite{2022Univ....8..547L}}

We notice that \cite{2022Univ....8..547L} reached a conclusion that the PWN's magnetic field is as high as $140 \mu$G, based on a model of the fast diffusion of the relativistic electrons injected into the PWN.
In the model, the steep profile of X-ray surface brightness is ascribed to the ballistic propagation of freshly injected electrons which occur only with a sufficient large diffusion coefficient. To reproduce the softening in the X-ray photon index profile, a strong magnetic field is needed to quickly cool these fast-diffusing electrons. As a result, the model in \cite{2022Univ....8..547L} requires a high magnetic field, which strongly suppress the IC radiation of electrons.

The model in this work takes into account a more realistic scenario for the PWN evolution and the particle transport. Therefore, we consider the spatial dependence of various key parameters, such as the magnetic field, the diffusion coefficient or the convection velocity. It leads to a distinct set of parameters in our model from that of \cite{2022Univ....8..547L}, and consequently different prediction of the UHE flux from electrons.

\subsection{Comparison with  \cite{2024ApJ...960...75P}}

\cite{2024ApJ...960...75P} also model the multiwavelength SED of the Boomerang nebula with the dynamical model. There are several differences in setups between their model and ours. First, in their favored model (Model C), they envisage a re-expansion phase of the PWN after being crushed by the SNR reverse shock, and limit the size of the PWN to be 100''. The magnetic field in their model is spatially homogeneous within the PWN. The measured X-ray profiles are not considered in the modeling. In our picture, the reverse shock has not arrived at the PWN and the latter is in the free-expansion phase. Also, we do not fix the outer boundary of the PWN to be 100''. Instead, we let the PWN evolve under the dynamical model, and use the measured X-ray profiles as a constraint. The second major difference is that \cite{2024ApJ...960...75P} employed the condition of $\eta_B+\eta_e=1$, i.e., the power of pulsar wind is converted to electrons and magnetic field, whereas in our model this condition is relaxed to $\eta_B+\eta_e\leq 1$ in light of some previous studies which suggest potential acceleration of protons in PWNe \cite{Atoyan1996, Amato2003, 2021Sci...373..425L, Liu2021}. Another difference is that \cite{2024ApJ...960...75P} assumed a broken power-law function for injected electron spectrum and fit the radio data while we assume a single power-law function and do not fit the radio data. As a result, their model does not work with a distance of 800\,pc and thus they favor a large distance of 7.5\,kpc for the nebula, based on the dispersion measurement of the pulsar and the NE2001 electron-density model \cite{NE2001}. Nevertheless, the obtained magnetic field of the PWN is of the order of $1-10\,\mu$G, which is consistent with our finding.



If the pulsar is located at $d=7.5\,$kpc distance, we cannot explain as previously the entire head region up to 900'' to be a part of the PWN, since otherwise it would result in an unusually large radius of 35\,pc for the PWN.  If future observations yield a preciser distance measurement of the pulsar's distance, the explanation of the X-ray profiles and LHAASO source need be revised.
In this case, the extended X-ray emission in the head region beyond 100'' may be explained as the escaping electrons/positrons from the nebula, similar to the previous suggestion for the very extended TeV nebula HESS~J1825-137 \citep{2020MNRAS.494.2618L}. Since a detailed investigation of the pulsar's distance is beyond the scope of this work, and $d=800$\,pc is employed as the distance of the pulsar in most of previous literature, we do not look further into the modeling under a larger pulsar's distance in this study.

\subsection{Electron acceleration efficiency}

\cite{2021Sci...373..425L} reported the detection of gamma rays from the Crab Nebula with the highest photon energy to be 1.1\,PeV. The observation indicated the presence of an electron PeVatron in the Crab nebula, with an extraordinary acceleration rate that need reach at least 15\% of the theoretical limit or $\epsilon >0.15$ in order to overcome the radiative loss. In all three scenarios studied in this work, the maximum electron energy is determined by confinement (Eq.~\ref{eq:gmax}), because of the small radius of the termination shock. As a result, the requirement of the acceleration efficiency $\epsilon$ \textbf{is higher than the value given by Crab} (see Table~\ref{tab:MCMC}). The maximum electron energy can reach a few PeV in the convection-dominated scenario, and approach 1\,PeV in the other two scenarios. Since the maximum particle energy is limited by the confinement, protons cannot reach a higher maximum energy than electrons. Nevertheless, without suffering from the KN suppression, the hadronic emission of protons accelerated by the PWN would still appear slightly harder than the IC emission at the UHE band, and make a contribution to the measured UHE emission. If the pulsar is located at a larger distance, the physical size of the termination shock would be correspondingly larger, the maximum energy of protons can be higher. 

\subsection{Proton fraction}

In our model, the condition of $\eta_B+\eta_e=1$ is relaxed. The best-fit results of X-ray data yield $\eta_B+\eta_e \ll 1$ for all three scenarios, leaving room for energisation of protons in the PWN.
One possible explanation is that over 90\% of the rotational energy of the Boomerang nebula could be converted into protons, which is significantly higher than the 1/3 proportion proposed by \cite{Amato2003}. LHAASO's data also constrain the Crab Nebula to convert 10-50\% of its spindown energy into relativistic protons, as reported by \cite{2021ApJ...922..221L}. 
While in our model, the Boomerang nebula has different physical conditions, such as magnetic field spatial configuration or asymmetric structure of the nebula, allowing for a higher proton energization. Therefore, our results suggest that the unique environment of the Boomerang Nebula is indeed capable of converting a larger proportion of energy into proton energy, which is not necessarily applicable to all PWN. 
Most of the spindown luminosity of the pulsar is transformed into energization of protons for the Boomerang nebula.

On the other hand, the rest of the spindown energy is not necessarily transferred to relativistic protons completely. Some of them could be in the form of thermal particles, as suggested by some particle-in-cell (PIC)  simulations of particle acceleration in relativistic shocks \citep{2009ApJ...698.1523S} or in relativistic reconnection  \citep{2020A&A...642A.204C}. The fraction of particle energy in the high-energy supra-thermal tails (i.e., with a power-law distribution) to that in the thermal bath depends on properties of the shock such as the magnetization and the magnetic obliquity. Therefore, the value of $(1 - \eta_e - \eta_B)$ may not be simply considered as the value of $\eta_p$ but rather an upper limit. \cite{2021Sci...373..425L} found the 10 PeV proton luminosity to be $(7-260)\times 10^{36}$\,erg/s, depending on the escape timescale of accelerated protons from the PWN. It corresponds to 1.5\% - 60\% of the spindown luminosity (if integrating over the entire spectrum, the value could be higher). The value has a large uncertainties and hence does not give a strong limit on the proton fraction.

\begin{figure}
    \centering
    \includegraphics[width=0.4\linewidth]{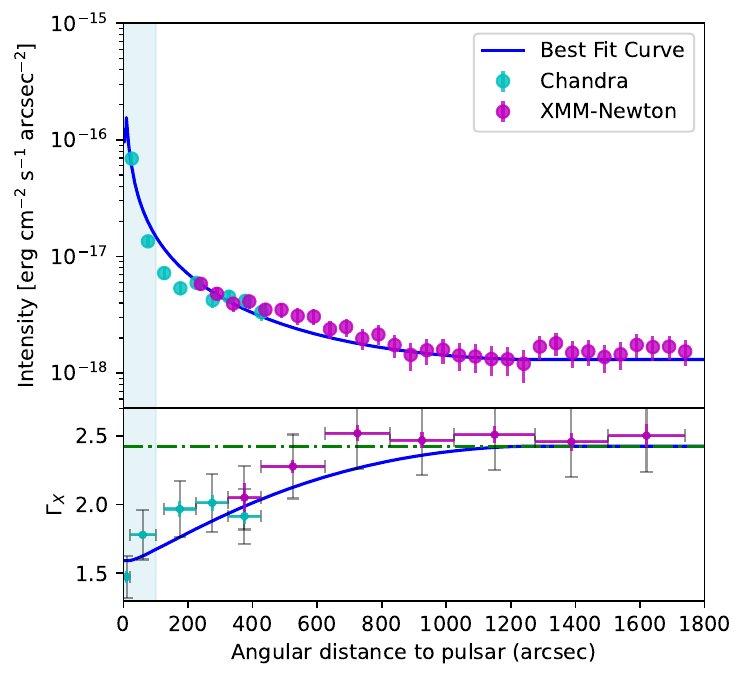}
    \caption{Fitting result of the radial profiles of X-ray surface brightness and photon index for forcing zero proton fraction. In this case, the parameters are as follows: $\eta_B =0.9972$, $\eta_e =2.818\times 10^{-3}$, $\alpha=1.91$, $\beta = 0.33$ and $\epsilon=0.52$. }
    \label{fig:etaB_eq_1}
\end{figure}

Finally, it may be also interesting to note that if we limit $\eta_B+\eta_e = 1$, the X-ray observations may also be roughly reproduced with our model (see Figure~\ref{fig:etaB_eq_1}. In this situation, we obtain $\eta_B (=0.9972)$ and $\eta_e (=2.818\times 10^{-3})$. It suggests the dominance of magnetic field energy at the TS, corresponding to $B_0(t=\rm now) = 98$\,$\mu$G. In this scenario, the Boomerang nebula is not expected to be a gamma-ray emitter, because the IC radiation is severely suppressed and we do not expect hadronic radiation neither.

\section{Conclusion} \label{sec:Conclu}
In conclusion, we modeled the X-ray emission of the Boomerang nebula including the intensity profile and the photon index profile. We employed the dynamical evolution model of PWN, and considered three scenarios for the electron transport, i.e., the convection-dominated scenario, the convection-diffusion scenario, and the diffusion-dominated scenario. The fitting results of all three scenarios converge into a relatively weak magnetic field of the PWN, with the strength $\sim 10\,\mu$G in the vicinity of the termination shock and decreasing to $\sim 1\,\mu$G, which is different from findings in some previous studies \citep{1941DoSSR..30..301K, 2006ApJ...638..225K}. With such a magnetic field, the IC radiation of electrons remains a significant radiative process within the PWN.
Depending on the particle transport scenario, the UHE gamma-ray flux of LHAASO~J2226+6057 contributed by the Boomerang nebula via the IC radiation may reach a level of about $10-50\%$ at 100\,TeV, and up to 30\% at 500\,TeV.
The rest of UHE gamma-ray flux may arise from SNR~G106.3+2.7, or hadronic emission of protons accelerated by the PWN. Indeed, in our model, the condition of $\eta_B+\eta_e=1$ is relaxed. The best-fit results of X-ray data yield $\eta_B+\eta_e \ll 1$ for all three scenarios, leaving a  room for energisation of protons in the PWN. In addition, the Boomerang nebula is surrounded by atomic gas in the northeast side, providing the target for hadronic interactions. 
In the future, more accurate TeV gamma-ray emissions, in particular the CTA observation with a high angular resolution may provide a TeV gamma-ray profile, which only depends on particle distribution (because the seed radiation field is homogeneous). This may help in determining the particle transport mechanism and distinguishing among the three cases.

\begin{acknowledgments}
This study is supported by National Natural Science Foundation of China under grants No.~12393852, 12333006, 12121003.
\end{acknowledgments}

\appendix
\section{MCMC fitting to the X-ray observations of the Boomerang nebula}
\begin{figure}
    \centering
    \includegraphics[width=0.5\linewidth]{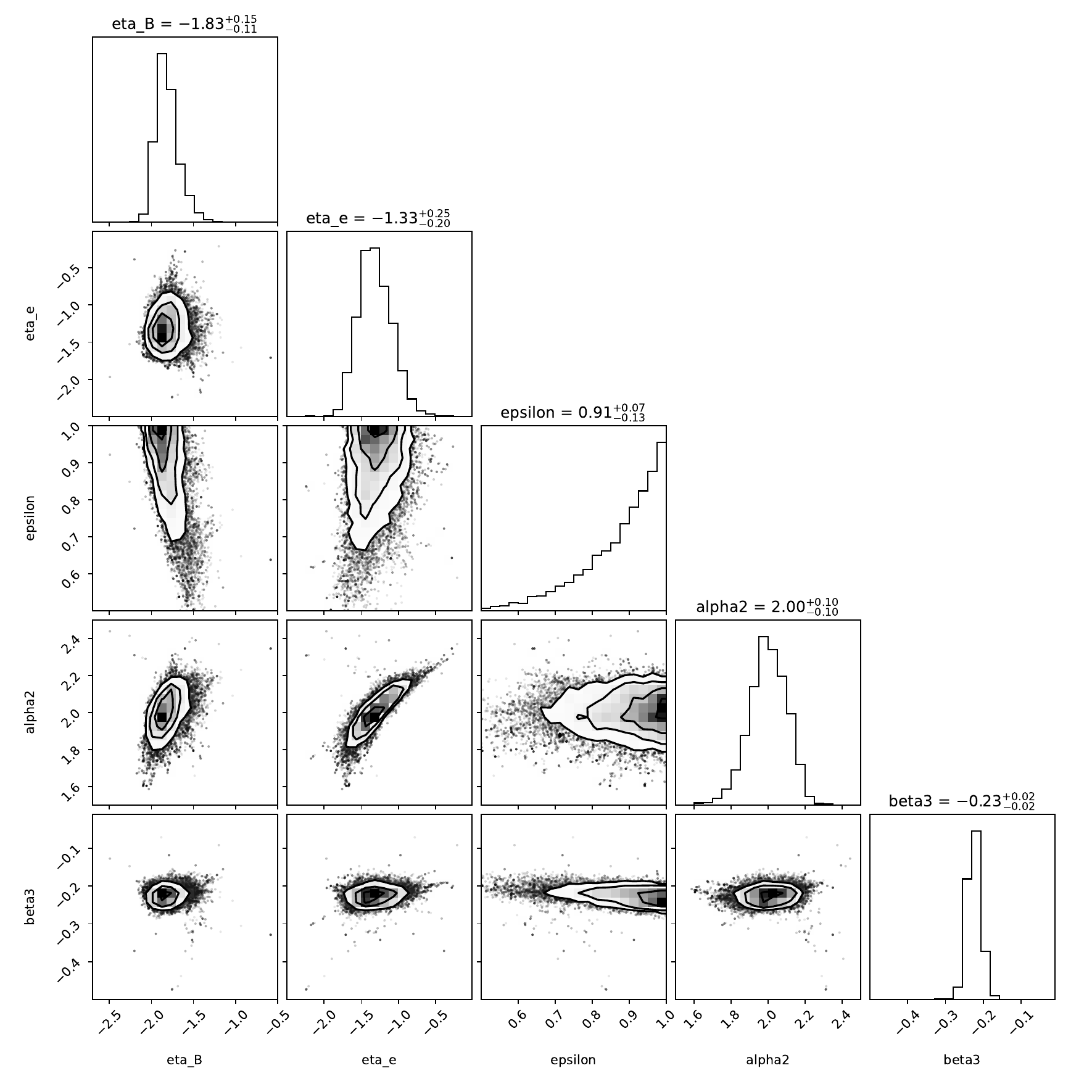}
    \caption{The corner plot of the MCMC fitting for the convection-dominated scenario.}
    \label{fig:contour1}
\end{figure}

\begin{figure}
    \centering
    \includegraphics[width=0.5\linewidth]{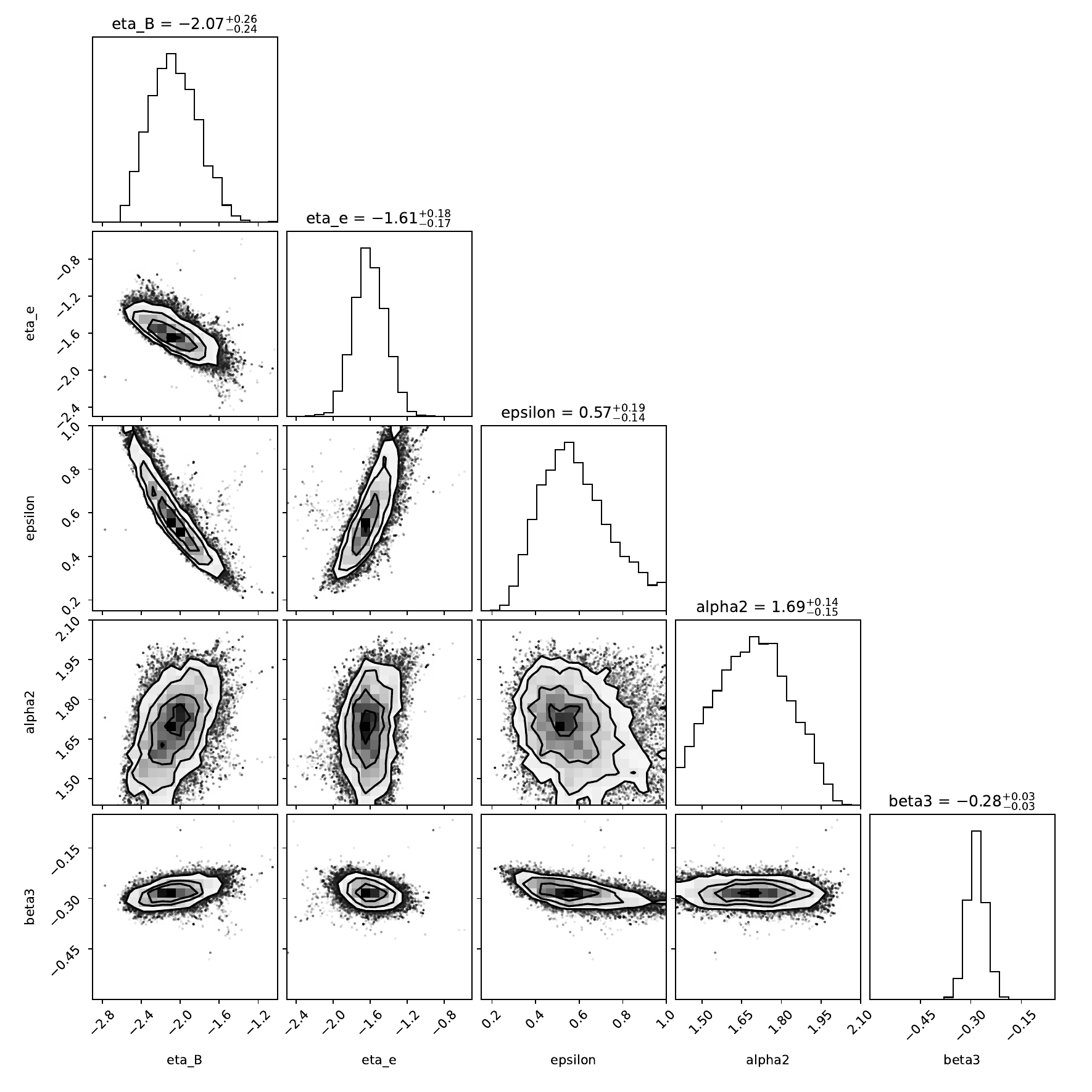}
    \caption{The corner plot of the MCMC fitting for the convection-diffusion scenario.}
    \label{fig:contour2}
\end{figure}

\begin{figure}
    \centering
    \includegraphics[width=0.5\linewidth]{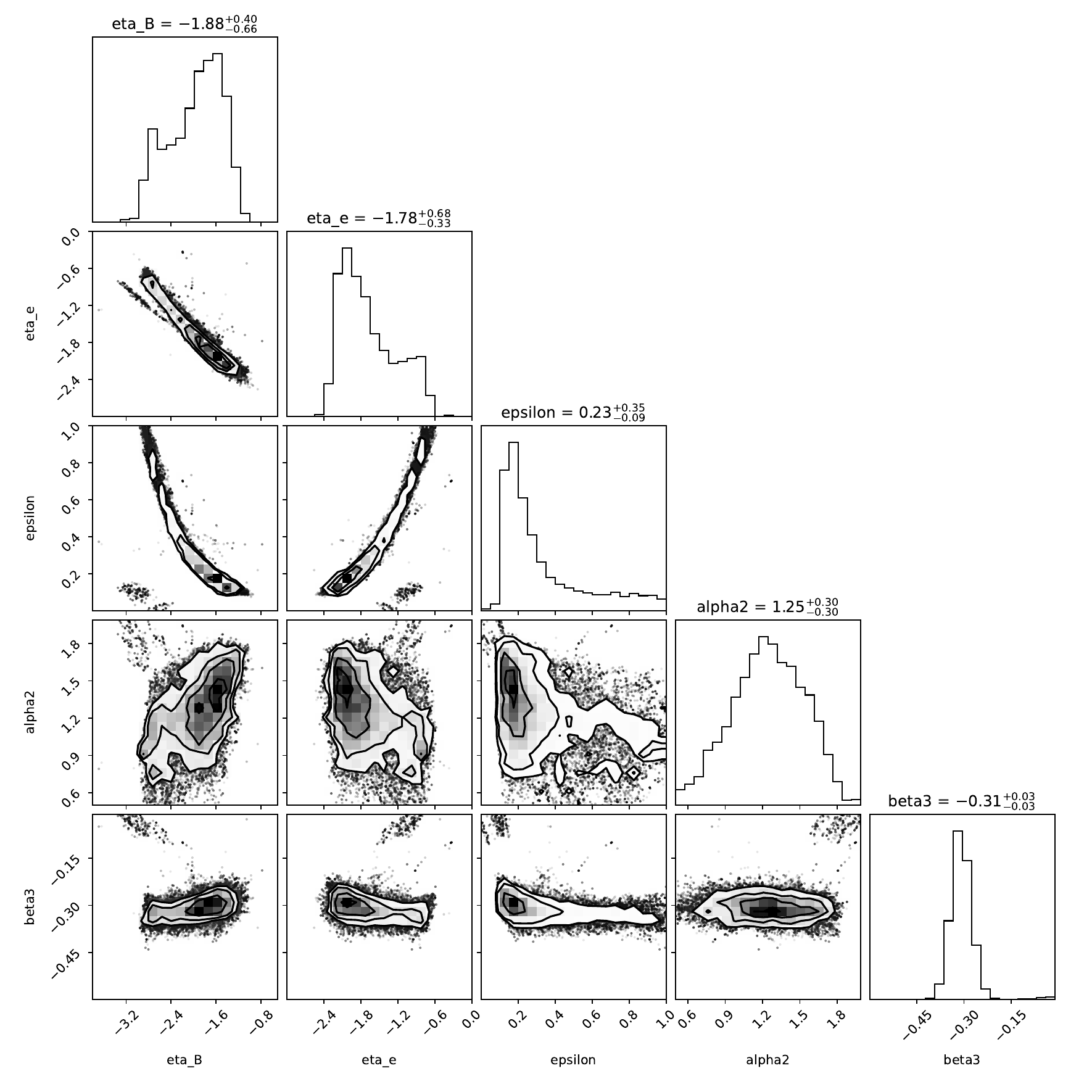}
    \caption{The corner plot of the MCMC fitting for the diffusion-dominated scenario.}
    \label{fig:contour3}
\end{figure}

The corner plots of the MCMC fitting to three electron transport scenarios are shown in Figures~\ref{fig:contour1}, \ref{fig:contour2} and \ref{fig:contour3}. 


\bibliography{sample631}{}
\bibliographystyle{aasjournal}

\end{document}